\documentclass[10pt,english]{article}
\usepackage{mathptmx}
\usepackage[T1]{fontenc}
\usepackage[latin9]{luainputenc}
\usepackage{babel}
\usepackage{amsmath}
\usepackage{amssymb}
\usepackage{graphicx}
\usepackage{setspace}
\usepackage[pdfusetitle,
 bookmarks=true,bookmarksnumbered=false,bookmarksopen=false,
 breaklinks=false,pdfborder={0 0 1},backref=false,colorlinks=false]
 {hyperref}

\makeatletter
\newcommand{\lyxaddress}[1]{
	\par {\raggedright #1
	\vspace{1.4em}
	\noindent\par}
}

\makeatother

\begin{document}
\title{A simplified Parisi Ansatz II:\\
Random Energy Model universality}
\author{Simone Franchini}
\date{~}
\maketitle

\lyxaddress{\begin{center}
{\small\textit{Sapienza Università di Roma, 1 Piazza Aldo Moro, 00185
Roma, Italy}}
\par\end{center}}
\begin{abstract}
In a previous work {[}\textit{A simplified Parisi Ansatz}, Franchini,
S., Commun. Theor. Phys., 73, 055601 (2021){]} we introduced a simple
method to compute the Random Overlap Structure of Aizenmann, Simm
and Stars and the full RSB Parisi formula for the Sherrington--Kirkpatrick
Model without using Replica Theory. The method consists in partitioning
the system into smaller sub-systems that we call layers, and iterate
the Bayes rule. A central ansatz in our derivation was that these
layers could be approximated by Random Energy Models of the Derrida
type. In this paper we analyze the properties of the interface in
detail, and show the equivalence with the Random Energy Model at any
temperature.

~

\end{abstract}
\pagebreak\tableofcontents{}\pagebreak{}

\section{Introduction}

The Sherrington--Kirkpatrick (SK) model is a well known toy model
for complex systems and plays a central role \cite{ParisiNobel} in
the celebrated ``Replica Symmetry Breaking'' (RSB) theory of spin
glasses (SG) by Parisi, Mezard, Virasoro \cite{Parisi} and many others
\cite{Charbonneau}.

This topic became of general interest in recent years: as explained
in the motivation of the Nobel Prize in Physics 2021 to Parisi \cite{NobelRel},
the RSB theory provided fundamental insights and applications in the
most distant fields, e.g., ranging from combinatorial optimization
\cite{Mezard_Montanari} to glasses, granular matter at jamming \cite{Benetti},
random lasers \cite{Ghirofania} and the problem of turbulence and
nonlinear wave interactions \cite{Gonzalez}. Of special importance
for the most recent scientific trends are the applications to neural
networks \cite{Amit}, both artificial and natural, and especially
those to understand the principles behind AI architectures \cite{Rende,Li_spomp,Tiberi}.
In general, it is expected that it may provide insights also in other
situations where nonlinear phenomena are involved, like stabilization
\cite{Stojanovic2}, fault estimation \cite{Stojanovic3} and integration
of PDE systems \cite{Stojanovic1}.

Although acclaimed and well known, the RSB theory is still quite technical
matter to deal with, especially when mathematical rigor is required.
To overcome these problems, in previous papers \cite{Franchini2021,Franchini2023,Franchini2024}
we introduced a method to study the SK model (and many other physical
systems) without replicas: as is shown in \cite{Franchini2021}, this
method allows a natural derivation of the ``Random Overlap Structure''
(ROSt) of Aizenmann, Simms and Starr \cite{ASS}, and the \textbf{full-RSB}
Parisi functional for computing the free energy per spin \cite{Franchini2021,Franchini2023}.
The main steps of our analysis where to introduce a sequence of SK
models of increasing sizes by partitioning the vertices set into subsets,
that we call layers, and then show that these layers can be approximated
by a simpler noise model, that we call \textbf{interface}, where the
Hamiltonian is simply the scalar product between the spin state and
some external field (see Lemma 10 of \cite{Franchini2023}). 

In \cite{Franchini2021}, a crucial claim was that the interfaces
can be approximated by Random Energy Models (REM) \cite{Kurkova1},
the simplest toy model for a disordered systems. Introduced by B.
Derrida in the 1980s, this model has inspired important mathematical
advances in the understanding of spin glasses, particularly trough
its relation with the Poisson Point Processes (PPP). Of special interest
is the \textbf{REM universality} \cite{Arous-Kupsov}. After several
precursor papers, worth to cite Ebeling and Nadler \cite{Nadler},
Mertens \cite{Mertens}, Borgs, Chayes and Pittel \cite{Borgs1},
etc., the REM universality has been finally recognized, by Mertens,
Franz and Bauke \cite{Numberpart}, in the context of combinatorial
optimization, and further investigated by other authors, see \cite{Arous-Kupsov}
for a survey. As is said in \cite{Arous-Kupsov}, the basic phenomenon
is in that the micro-canonical distribution of the energies of a large
class of models is close to a REM in distribution for certain energy
windows. 

By implementing a form of REM universality, in the Lemma 12 of \cite{Franchini2023}
we showed that the thermal fluctuations of the interface near the
ground state converge to a REM at near zero temperature. In this paper
we study the interface model in detail, compute the thermodynamic
limit, and show the equivalence with the REM at any temperature in
case of Gaussian noise. We remark that the present paper aims to describe
the interface model, i.e. the \textbf{one body Hamiltonian} of Eq.
(\ref{eq:noisemodel}) below, that to best of our knowledge does not
have a dedicated paper describing its properties. We will not discuss
the SK model here, although the results can be obviously applied to
the SK model following the methods shown in \cite{Franchini2021,Franchini2023,Franchini2024}. 

\section{Summary}

Let briefly introduce the basic notation. Let $V=\left\{ 1,2,\,...\,,N\right\} $
be a set of $N$ vertices and put a spin $\sigma_{i}\in\Omega$ of
inner states $\Omega=\{+,-\}$ on each vertex, we denote by 
\begin{equation}
\sigma_{V}:=\left\{ \sigma_{i}\in\Omega:\,i\in V\right\} 
\end{equation}
the generic magnetization state. The support of $\sigma_{V}$ is the
product space $\Omega^{V}$. We denote by $\mathbb{I}\left(A\right)$
the indicator function of the event $A$, that is $\mathbb{I}\left(A\right)=1$
if $A$ is verified and zero otherwise. Also, given two ordered sets
$A$ and $B$ we use notation $A\otimes B$ for the tensor product
and just $A\,B$ for the Cartesian product. The scalar product is
denoted by the usual $\cdot$ symbol. The Hadamard product is denoted
by the $\circ$ symbol. 

\subsection{The interface model}

Following ideas from Borgs, Chayes \cite{Borgs2,Borgs3}, Coja-Oghlan
\cite{Coja-Oghlan} and others, in a recent paper we showed \cite{Franchini2023}
that the scalar product of the spin state with some external field,
that we call \textbf{interface model}, and formally describe with
the Hamiltonian
\begin{equation}
H\left(\sigma_{V}|h_{V}\right):=\sigma_{V}\cdot h_{V},\label{eq:noisemodel}
\end{equation}
can be approximated by the REM in the low temperature phase. The field
components are real numbers $h_{i}\in\mathbb{R}$ indexed by $i\in V$
and are assumed to have been independently extracted from some probability
distribution $p$. We use a braket notation for the average of the
test function $f$ respect to the Gibbs measure $\xi$ (\textbf{softmax})
\begin{equation}
\langle f\left(\sigma_{V}\right)\rangle_{\xi}:=\sum_{\sigma_{V}\in\Omega^{V}}f\left(\sigma_{V}\right)\,\exp\left[-\sigma_{V}\cdot\beta h_{V}+N\beta\zeta\left(\beta h_{V}\right)\right],\label{eq:average}
\end{equation}
where $N$ is the number of spins and $\zeta$ is the free energy
density per spin:
\begin{equation}
-\beta\zeta\left(\beta h_{V}\right):=\frac{1}{N}\sum_{i\in V}\log2\cosh\left(\beta h_{i}\right).\label{eq:scalinglimit-1}
\end{equation}
We remark that the interface model is closely related to the ``Number
Partitioning Problem'' (NPP) \cite{Numberpart,Borgs2,Borgs3}, and
we may also refer to it as a random field model, or noise model. It
correspond for example to the Random Field Ising Model (RIFM) in the
limit of zero Ising interactions (or infinite field amplitude) and
many other models. In general, the interface could be seen as the
zero interaction limit of any lattice field theory of the kind described
in \cite{Bardella_Franchini}.

\subsection{Thermodynamic limit }

In Section \ref{subsec:Field-fluctuations} we study the thermodynamic
limit by quantile mechanics \cite{Quantile=000020mechanics} and series
analysis, and give explicit examples for the binary, uniform and Gaussian
cases. The scaling limit of the \textbf{Free Energy density} for an
infinite number of spins will converge almost surely to the following
functional:
\begin{equation}
-\lim_{N\rightarrow\infty}\beta\zeta\left(\beta h_{V}\right)=\int_{0}^{1}dq\,\log2\cosh\left[\beta x\left(q\right)\right].\label{eq:scalinglimit}
\end{equation}
The function $x\left(q\right)$ is called \textbf{quantile} and is
found by inverting the cumulant of $p$,
\begin{equation}
q\left(x\right):=\int_{0}^{x}dz\ p\left(z\right),
\end{equation}
by quantile mechanics \cite{Quantile=000020mechanics} the quantile
satisfies the differential equation
\begin{equation}
\partial_{q}^{2}x\left(q\right)=\rho\left[\,x\left(q\right)\right]\left[\partial_{q}\,x\left(q\right)\right]^{2},
\end{equation}
where the function $\rho$ is defined from $p$ according to the relation
\begin{equation}
\rho\left(x\right):=-\partial_{x}\log p\left(x\right).
\end{equation}
We explicitly compute the uniform and Gaussian cases. At high temperature
we find that, as expected, the free energy is replica symmetric, and
is therefore linear in temperature. At low temperature we find that
the convergence toward the ground state energy is quadratic in the
temperature, ie., the specific heat is linear like in the \textbf{Dulong-Petit}
law. The origin of this quadratic convergence is due to vertices with
small field amplitude (see Section \ref{sec:Thermodynamic-limit}),
and notice that, if we restrict to linear terms, the low temperature
modes can be neglected and the free energy is approximately constant
in temperature, like in the REM. In Section 5 of \cite{Franchini2023}
the convergence to the REM is actually shown in distribution in the
near zero temperature phase. 

\subsection{REM universality}

To this scope we introduce the \textbf{eigenstates of magnetization}
\begin{equation}
\Omega\left(m\right):=\left\{ \sigma_{V}\in\Omega^{V}:\,M(\sigma_{V})=\left\lfloor mN\right\rfloor \right\} ,
\end{equation}
where $M(\sigma_{V})$ stands for the total magnetization of the state
$\sigma_{V}$. These central objects of our analysis are studied in
detail in Section \ref{sec:Eigenstates-of-magnetization}. We also
introduce the ``master direction'' $\omega_{V}$, the flickering
state $\sigma_{V}^{*}$ and flickering function $f^{*}$
\begin{equation}
\omega_{i}:=h_{i}/|h_{i}|,\ \ \ \sigma_{i}^{*}:=\sigma_{i}\,\omega_{i},\ \ \ f^{*}\left(\sigma_{V}\right):=f\left(\sigma_{V}\circ\omega_{V}\right),
\end{equation}
where $\omega_{i}$ is the direction of the ground state in the vertex
$i$, and $f$ is any test function if not specified otherwise. We
can now introduce a fundamental variable, that we call $J-$field:
let define the following quantities
\begin{equation}
\psi:=\frac{1}{N}\sum_{i\in V}|h_{i}|,\ \ \ \delta^{2}:=\frac{1}{N}\sum_{i\in V}h_{i}^{2}-\psi^{2},\ \ \ J_{i}:=\frac{|h_{i}|-\psi}{\delta},
\end{equation}
where $\psi$ denotes the ground state energy density and $\delta$
is the amplitude of the fluctuations of $|h_{i}|$ around $\psi$.
As explained in the Sections \ref{sec:Binary-noise-model} and \ref{sec:Relation-with-the},
it is possible to track the fluctuations around the ground state.
This is done by introducing the the vertex set $X$, 
\begin{equation}
X\left(\sigma_{V}^{*}\right):=\left\{ i\in V:\,\sigma_{i}^{*}=-1\right\} ,
\end{equation}
that collects the vertices in which the spin $\sigma_{i}$ is flipped
with respect to the direction of the ground state $\omega_{i}$, and
the renormalized field $J$: 
\begin{equation}
J(\sigma_{V}^{*}):=\frac{1}{\sqrt{|X(\sigma_{V}^{*})|}}\sum_{i\in X(\sigma_{V}^{*})}J_{i},
\end{equation}
that is the normalized sum of the flipped local fields. The Hamiltonian
of Eq. (\ref{eq:noisemodel}) can be expressed in terms of $M$, $J$
and $X$ as follows:
\begin{equation}
\sigma_{V}\cdot h_{V}=\psi M\left(\sigma_{V}^{*}\right)+J(\sigma_{V}^{*}){\textstyle \sqrt{4\delta^{2}|X(\sigma_{V}^{*})|}}.
\end{equation}
In Section \ref{sec:Relation-with-the} we show that in the thermodynamic
limit the average of Eq. (\ref{eq:average}) is mostly sampled from
eigenstates of magnetization with eigenvalue that is found inverting\footnote{In the published version of this paper {[}Chaos, Solitons \& Fractals
191 (2025) 115821{]} is wrongly stated that the shape of $m_{0}$
is the hyperbolic tangent but this is true only in the low temperature
limit. }
\begin{equation}
\tanh^{-1}\left(m_{0}\right)+\frac{\beta^{2}\delta^{2}}{2}\,m_{0}=-\beta\psi,\ \ \ m_{0}^{2}\geq m_{c}^{2}\label{eq:emmezero}
\end{equation}
when $m_{0}^{2}$ is above some critical surface $m_{c}^{2}$, and
by inverting
\begin{equation}
\tanh^{-1}\left(m_{0}\right)\sqrt{\frac{1-m_{0}^{2}}{\phi\left(m_{0}\right)}}+2m_{0}\,\sqrt{\frac{\phi\left(m_{0}\right)}{1-m_{0}^{2}}}=-\frac{2\psi}{\delta},\ \ \ m_{0}^{2}<m_{c}^{2}\label{eq:emezero2}
\end{equation}
if otherwise is below. 
\begin{equation}
\phi\left(m\right):=\log2-\frac{1}{2}\log\left(1-m^{2}\right)+\frac{m}{2}\log\left(\frac{1-m}{1+m}\right)
\end{equation}
is the binary entropy. The equation 
\begin{equation}
\phi\left(m_{c}\right)=\frac{\beta^{2}\delta^{2}}{4}{\textstyle \left(1-m_{c}^{2}\right)}\label{eq:emmezero3}
\end{equation}
identifies the critical surface $m_{c}$, where the two branches connect.
For any $\sigma_{V}\in\Omega\left(m_{0}\right)$ we can rewrite the
Hamiltonian as:
\begin{equation}
\sigma_{V}\cdot h_{V}=\psi\,m_{0}\,N+J(\sigma_{V}^{*}){\textstyle \sqrt{K\left(m_{0}\right)}}\label{eq:zanza}
\end{equation}
where we introduced the auxiliary function 
\begin{equation}
K(m):=2\delta^{2}(1-m)\,N.\label{eq:K_min}
\end{equation}
In reference \cite{Franchini2023} we found that at low temperature
the $J$ field applied to $\Omega\left(m_{0}\right)$ actually converges
in distribution to a REM. This is done by noticing that when the temperature
goes to zero the state align toward the direction of the ground state
almost everywhere, and only a small fraction of spins is flipped in
the opposite direction. Since the flipped spins are sparse two independent
spin configurations will probabily have a small number of common flipped
spins, that can be ignored, making the corresponding $J-$fields independent.
In Section \ref{sec:Relation-with-the} we show that it is possible
to extend the results of \cite{Franchini2023} to the full temperature
range by properly renormalizing the $J$ field. In particular, we
will introduce the $J'$ field, described in detail in the Section
\ref{sec:Relation-with-the}. It is shown that when applied to the
ensemble $\Omega(m_{0})$, this field is distributed like a REM by
construction (i.e., a field where the pairwise overlap matrix is zero
on average). In Section \ref{sec:Relation-with-the} we show that
the $J$ is distributed like $J'$ up to a constant and a renormalization
\begin{equation}
J(\sigma_{V}){\textstyle \sqrt{K(m_{0})}}=J'(\sigma_{V}){\textstyle \sqrt{K'(m_{0})}}+\mathrm{const.},\ \ \ K'(m):=\delta^{2}(1-m^{2})\,N,\label{eq:fianeq}
\end{equation}
the renormalized amplitude is found in Section \ref{sec:Relation-with-the}.
By the averaging properties of PPP, a parameter $\lambda$ exists
(dependent from $\beta$, $\psi$ and $\delta$) such that the Gibbs
average satisfy the REM-PPP average formula \cite{Franchini2021,Franchini2023,ASS,Kurkova1},
\begin{equation}
\lim_{N\rightarrow\infty}\langle f\left(\sigma_{V}\right)\rangle_{\xi}=\lim_{N\rightarrow\infty}\langle f^{*}\left(\sigma_{V}\right)^{\lambda}\rangle_{m_{0}}^{1/\lambda},\label{eq:REM-PPP}
\end{equation}
that in the subcritical region $\lambda\in\left[0,1\right]$ interpolates
between the geometric and the arithmetic average. The importance of
the REM universality to the spin glass physics is now evident in that
one could directly apply this formula to Eq. (55) of \cite{Franchini2021}
(or Eq. (6.20) of \cite{Franchini2023}) and find the \textbf{full-RSB}
Parisi functional. This completes the steps to compute the free energy
of the SK model (and many other models) with methods and concepts
from \cite{Franchini2021,Franchini2023,Franchini2024}, see Section
\ref{sec:Relation-with-the} below and Section 5 of \cite{Franchini2023}
for further details. Notice that, apart from disordered systems and
lattice field theory, similar properties have been recently observed
also in important neural network models. Of special interest is the
connection with \textbf{Dense Associative Memories} (DAM): for example,
in \cite{Mezard_Lucibello} has been shown that also in the exponential
Hopfield models one can approximate the free energy of each layer
with that of a REM. 

\section{Thermodynamic limit \protect\label{sec:Thermodynamic-limit}}

Let us start by formally defining the interface as in \cite{Franchini2021,Franchini2023}:
the Hamiltonian is that of Eq. (\ref{eq:noisemodel}). Following the
canonical notation, we called $\beta$ the inverse of the temperature.
The canonical partition function is defined as follows:
\begin{equation}
Z\left(\beta h_{V}\right):=\sum_{\sigma_{V}\in\Omega^{V}}\exp\left(-\sigma_{V}\cdot\beta h_{V}\right),
\end{equation}
the associated Gibbs measure is given by the expression
\begin{equation}
\xi\left(\sigma_{V}|\beta h_{V}\right):=\frac{\exp\left(-\sigma_{V}\cdot\beta h_{V}\right)}{Z\left(\beta h_{V}\right)}=\exp\left[-\sigma_{V}\cdot\beta h_{V}+N\beta\zeta\left(\beta h_{V}\right)\right]
\end{equation}
where the function $\zeta$ is the free energy density per spin
\begin{equation}
-\beta\zeta\left(\beta h_{V}\right):=\frac{1}{N}\log Z\left(\beta h_{V}\right)=\frac{1}{N}\sum_{i\in V}\log2\cosh\left(\beta h_{i}\right).
\end{equation}
Let $f$ be a test function of $\sigma_{V}$ the Gibbs average is
as follows
\begin{equation}
\langle f\left(\sigma_{V}\right)\rangle_{\xi}:=\sum_{\sigma_{V}\in\Omega^{V}}f\left(\sigma_{V}\right)\,\exp\left[-\sigma_{V}\cdot\beta h_{V}+N\beta\zeta\left(\beta h_{V}\right)\right].
\end{equation}

\subsection{Field fluctuations\protect\label{subsec:Field-fluctuations}}

Since the free energy depends only on the absolute value of the external
field and not on the direction, before proceeding with the computations
it is convenient to introduce the following auxiliary variables:
\begin{equation}
x_{i}=\left|h_{i}\right|,\ \ \ \omega_{i}=h_{i}/|h_{i}|,\ \ \ \sigma_{i}^{*}=\sigma_{i}\,\omega_{i}
\end{equation}
the flickering state $\sigma_{V}^{*}$ is the Hadamard product between
the initial state $\sigma_{V}$ and the direction of the external
field $\omega_{V}$, that corresponds to the ground state of the system
and that we call master direction. Using these variables the Hamiltonian
can be rewritten as $\sigma_{V}^{*}\cdot x_{V}$ where $x_{V}$ is
a vector with all positive entries, we call it rectified field. To
find the scaling limit of the free energy it is convenient to reoreder
$V$ according to the \textbf{order statistics}, a remapping of the
index $i\in V$, usually denoted with the symbol $\left(i\right)$,
such that $x_{\left(i\right)+1}\geq x_{\left(i\right)}$. Then, it
is easy to see that if each $x_{i}$ is independently drawn according
to the same probability density $p$, then the scaling limit of $x_{\left(i\right)}$
for $\left(i\right)/N\rightarrow q\in\left[0,1\right]$ will almost
surely converge to the quantile function of $p$. 

\subsection{Uniform distribution\protect\label{subsec:Uniform-distribution}}

There are several important cases that can be treated exactly, one
could consider the uniform distribution $p\left(x\right)=\mathbb{I}\left(x\in\left[0,1\right]\right)$:
here the cumulant is $q\left(x\right)=x$ and the quantile is therefore
$x\left(q\right)=q$, then, the free energy density converges to the
following integral:
\begin{equation}
-\lim_{N\rightarrow\infty}\beta\zeta\left(\beta x_{V}\right)=\int_{0}^{1}dq\,\log2\cosh\left(\beta q\right)=\frac{1}{\beta}\int_{0}^{\beta}d\tau\,\log2\cosh\left(\tau\right)
\end{equation}
the primitive of this integral is easily found via computer algebra,
\begin{equation}
\int d\tau\,\log2\cosh\left(\tau\right)=\frac{\mathrm{Li_{2}\left[-\exp\left(-2\tau\right)\right]}+\tau^{2}}{2}
\end{equation}
where $\mathrm{Li_{2}}$ is the dilogarithm \cite{Zaiger}, or Spence's
function, that is often encountered in particle physics: for $z\in\left[0,1\right]$
the following relations holds 
\begin{equation}
\mathrm{Li_{2}}\left(-z\right)=\sum_{k\geq1}\frac{\left(-z\right)^{k}}{k^{2}},\ \ \ \mathrm{Li_{2}\left(-1\right)}=-\frac{\pi^{2}}{12},\ \ \ \mathrm{Li_{2}}\left(0\right)=0,
\end{equation}
the first derivative obeys the following formula
\begin{equation}
\partial_{z}\,\mathrm{Li_{2}}\left(-z\right)=\sum_{k\geq1}\frac{\left(-z\right)^{k-1}\left(-1\right)^{k}}{k}=z^{-1}\sum_{k\geq1}\frac{\left(-z\right)^{k}}{k}=-\frac{\log\left(1+z\right)}{z}
\end{equation}
and notice that at $-1$ the derivative converges to the nontrivial
value
\begin{equation}
\partial_{z}\mathrm{Li_{2}}\left(-1\right)=-\log2
\end{equation}
Then, the scaling limit of the free energy density converges to the
expression
\begin{equation}
-\lim_{N\rightarrow\infty}\beta\zeta\left(\beta x_{V}\right)=\frac{\beta}{2}+\frac{\pi^{2}}{24\,\beta}+\frac{\mathrm{Li_{2}\left[-\exp\left(-2\beta\right)\right]}}{2\beta},
\end{equation}
where the last term is negative in the whole temperature range. Notice
that the convergence to the ground state energy is quadratic in temperature:
the specific heat is linear like in the Dulong-Petit law.

\subsection{Half-Gaussian distribution\protect\label{subsec:Half-gaussian-distribution}}

We could also consider more complex shapes, like the half-normal distribution,
\begin{equation}
p\left(x\right)=\sqrt{2/\pi}\ \exp\left(-x^{2}/2\right).
\end{equation}
From quantile mechanics \cite{Quantile=000020mechanics} one finds
\begin{equation}
\rho\left(x\right)=-\partial_{x}(-x^{2}/2+\log\sqrt{2/\pi}\,)=x,
\end{equation}
then, the quantile equation and its boundary conditions is as follows:
\begin{equation}
\partial_{q}^{2}x\left(q\right)=x\left(q\right)\left[\partial_{q}\,x\left(q\right)\right]^{2},\ \ \ x\left(0\right)=0,\ \ \ x\left(1/2\right)=\sqrt{2}\ \mathrm{erf^{-1}}\left(1/2\right),\label{eq:gaussy}
\end{equation}
solving the equation with these boundary give us 
\begin{equation}
x\left(q\right)=\sqrt{2}\ \mathrm{erf^{-1}}\left(q\right).
\end{equation}
Then, for the half-normal (and normal) noise model we expect the free
energy to converge toward the following limit expression 
\begin{equation}
-\lim_{N\rightarrow\infty}\beta\zeta\left(\beta x_{V}\right)=\int_{0}^{1}dq\,\log2\cosh\,[\,\beta\sqrt{2}\ \mathrm{erf^{-1}}\left(q\right)].
\end{equation}
We notice that the differential equation in Eq. (\ref{eq:gaussy})
before is remarkably similar to the term in parentesis in Eq. (10)
of \cite{Guerra}, that is the nonlinear antiparabolic equation from
which we obtain the Parisi functional. Further investigation on the
relation between the Guerra interpolation theory and quantile mechanics
would be certainly interesting, we hope to explore this in a future
work.

\subsection{Series analysis of the Half-Gaussian}

To highligt the low temperature features it will be more instructive
to rather perform a series analysis. We start from the expression
\begin{equation}
-\lim_{N\rightarrow\infty}\beta\zeta\left(\beta x_{V}\right)=\lim_{N\rightarrow\infty}\frac{1}{N}\sum_{i\in V}\log2\cosh\left(\beta x_{i}\right),
\end{equation}
the ground state in the thermodynamic limit (TL) is
\begin{equation}
-\lim_{N\rightarrow\infty}\zeta\left(\infty\right)=\lim_{N\rightarrow\infty}\lim_{\beta\rightarrow\infty}\frac{1}{N}\sum_{i\in V}x_{i}\tanh\left(\beta x_{i}\right)=\lim_{N\rightarrow\infty}\frac{1}{N}\sum_{i\in V}x_{i}=\sqrt{\frac{2}{\pi}},
\end{equation}
where the numerical value was obtained by computing the average of
$x$ with $x$ Gaussian variable of zero average and unitary variance:
\begin{equation}
\lim_{N\rightarrow\infty}\frac{1}{N}\sum_{i\in V}x_{i}=\sqrt{\frac{2}{\pi}}\int_{0}^{\infty}x\,\exp\left(-x^{2}/2\right)\,dx=\sqrt{\frac{2}{\pi}}\int_{0}^{\infty}\exp\left(-t\right)\,dt=\sqrt{\frac{2}{\pi}},
\end{equation}
in the last step we applied the substitution $t=x^{2}/2$. Now, let
us consider the equivalent expression
\begin{equation}
\lim_{N\rightarrow\infty}\frac{1}{N}\sum_{i\in V}\log2\cosh\left(\beta x_{i}\right)=\beta\sqrt{\frac{2}{\pi}}+\lim_{N\rightarrow\infty}\frac{1}{N}\sum_{i\in V}\log\left[1+\exp\left(-2\beta x_{i}\right)\right],
\end{equation}
the logarithm can be expanded in the limit of large $\beta$,
\begin{equation}
\log\left[1+\exp\left(-2k\beta x_{i}\right)\right]=\sum_{k\geq1}\frac{\left(-1\right)^{k-1}}{k}\exp\left(-2k\beta x_{i}\right),
\end{equation}
the average of the exponential term can be computed by Gaussian integration
\begin{multline}
\lim_{N\rightarrow\infty}\frac{1}{N}\sum_{i\in V}\exp\,(-2k\beta x_{i})=\sqrt{\frac{2}{\pi}}\int_{0}^{\infty}dx\,\exp\,(-2k\beta x-x^{2}/2)=\\
=\exp\,(\,2k^{2}\beta^{2})\,\sqrt{\frac{2}{\pi}}\int_{0}^{\infty}dx\,\exp\,[-(\sqrt{2}k\beta+x/\sqrt{2}\,)^{2}]=\\
=\exp\,(\,2k^{2}\beta^{2})\,\sqrt{\frac{4}{\pi}}\int_{0}^{\infty}d(\sqrt{2}k\beta+x/\sqrt{2}\,)\,\exp\,[-(\sqrt{2}k\beta+x/\sqrt{2}\,)^{2}]=\\
=\exp\,(\,2k^{2}\beta^{2})\,\sqrt{\frac{4}{\pi}}\int_{\sqrt{2}k\beta}^{\infty}dt\,\exp\,(-t^{2})=\exp\,(\,2k^{2}\beta^{2})\ \mathrm{erfc}\,(\sqrt{2}k\beta),\label{eq:bhjh}
\end{multline}
the last substitution is $t=x/\sqrt{2}+\sqrt{2}k\beta$. We found
a series representation for the free energy of the Gaussian noise
model
\begin{equation}
-\lim_{N\rightarrow\infty}\beta\zeta\left(\beta x_{V}\right)=\beta\sqrt{\frac{2}{\pi}}+\sum_{k\geq1}\frac{\left(-1\right)^{k-1}}{k}\exp\,(2k^{2}\beta^{2})\,\mathrm{erfc}(\sqrt{2}k\beta).
\end{equation}
For large $\beta$ one can use the following approximation for the
Error Function 
\begin{equation}
\exp\,(2k^{2}\beta^{2})\,\mathrm{erfc}(\sqrt{2}k\beta)=\frac{1}{\sqrt{2\pi}k\beta}+O\left(\beta^{-3}\right).
\end{equation}
Put the last expression back into the logarithm expansion before and
one finds:
\begin{equation}
\lim_{N\rightarrow\infty}\frac{1}{N}\sum_{i\in V}\log\left[1+\exp\left(-2\beta x_{i}\right)\right]=\frac{K_{0}}{\sqrt{2\pi}\beta}+O\left(\beta^{-3}\right),\label{eq:thelimit}
\end{equation}
where the constant $K_{0}$ is the convergent sum 
\begin{equation}
K_{0}:=\sum_{k\geq1}\frac{\left(-1\right)^{k-1}}{k^{2}}=\frac{\pi^{2}}{12}
\end{equation}
The asymptotic expansion of $\zeta$ near zero temperature is then
found to be
\begin{equation}
-\lim_{N\rightarrow\infty}\beta\zeta\left(\beta x_{V}\right)=\beta\sqrt{\frac{2}{\pi}}+\frac{1}{\beta}\sqrt{\frac{\ \pi^{3}}{48}}+O\left(\beta^{-3}\right),
\end{equation}
also in this case the convergence to the ground state is quadratic
in temperature. 

\subsection{Convergence to the ground state\protect\label{subsec:Convergence-to-the}}

The origin of the quadratic difference in the low temperature behavior
is in that for both uniform and Gaussian distributions some couplings
may be close to zero for a fraction of spins. To see this, let analyze
the energy contribution from the subset of spins 
\begin{equation}
V\left(\delta\right)=\left\{ i\in V:\:x_{i}>\delta\right\} ,
\end{equation}
following the steps before we find 
\begin{multline}
\int_{\delta}^{\infty}dx\,\exp\,(-2k\beta x-x^{2}/2)=\\
=\exp\,(2k^{2}\beta^{2})\,\mathrm{erfc}(\delta/\sqrt{2}+\sqrt{2}k\beta)=O\left[\exp\,(-2\delta k\beta)\right].\label{eq:gggggg}
\end{multline}
and then the limit in Eq. (\ref{eq:thelimit}) restricted to $V\left(\delta\right)$
is 
\begin{equation}
\lim_{N\rightarrow\infty}\ \frac{1}{|V\left(\delta\right)|}\sum_{i\in V\left(\delta\right)}\log\left[1+\exp\left(-2\beta x_{i}\right)\right]=O\left[\exp\,(-2\delta\beta)\right],
\end{equation}
we can see if we exclude the spins with small coupling the temperature
dependence is again exponentially suppressed in $\beta$. 

\section{Eigenstates of magnetization\protect\label{sec:Eigenstates-of-magnetization}}

We introduce a central element in our analysis: the kernel of the
eigenstates of magnetization. Let define the microcanonical set
\begin{equation}
\Omega\left(M\right):=\left\{ \sigma_{V}\in\Omega^{V}:\,M\left(\sigma_{V}\right)=M\right\} ,\label{eq:MAGKERNEL}
\end{equation}
that is an ensemble collecting the magnetization states with a given
eigenstate $M$, or a lattice gas with exactly $E$ particles. Hereafter
we denote by $|\Omega\left(M\right)|$ the number of such eigenstates,
that we will call cardinality of $\Omega\left(M\right)$, or ``complexity'',
as is sometimes found in the spin glass literature. Also, define a
notation for the average, 
\begin{equation}
\langle f\left(\sigma_{V}\right)\rangle_{\Omega\left(M\right)}:=\frac{1}{|\Omega\left(M\right)|}\sum_{\sigma_{V}\in\Omega\left(M\right)}f\left(\sigma_{V}\right).
\end{equation}
The eigenstates of magnetization $\Omega\left(M\right)$ can be studied
in detail using Large Deviation Theory (LDT) \cite{Dembo_Zeitouni},
even at the ``sample-path'' LDT level. For example, by a simple
applications of the Varadhan Integral Lemma, the Mogulskii theorem,
and other standard LDT methods \cite{Dembo_Zeitouni,Franchini_Thesis,Franchini_URNS,Franchini_IRT,Franchini_RW2,Jack},
one can compute the number of the eigenstates with given magnetization
$\left\lfloor mN\right\rfloor $, ie the integer part of $mN$, with
$m\in\left[-1,1\right]$: to simplify the notation, hereafter
\begin{equation}
\Omega\left(\left\lfloor mN\right\rfloor \right)=:\Omega\left(m\right),\ \ \ \langle f\left(\sigma_{V}\right)\rangle_{\Omega\left(\left\lfloor mN\right\rfloor \right)}=:\langle f\left(\sigma_{V}\right)\rangle_{m}.
\end{equation}
With some algebraic effort is possible to show that the cardinality
of $\Omega\left(m\right)$ is proportional to $\exp\left[\,N\phi\left(m\right)\right]$,
with rate function $\phi$ equal to 
\begin{equation}
\phi\left(m\right)=\log2-\frac{1}{2}\log\left(1-m^{2}\right)+\frac{m}{2}\log\left(\frac{1-m}{1+m}\right),\label{eq:eigenstatesrate}
\end{equation}
this result can be obtained by applying the inverse Legendre transform
to the free energy of the binary noise model, studied in Section \ref{sec:Binary-noise-model}.
In Ref. \cite{Franchini_Thesis,Franchini_URNS} a detailed description
of the magnetization eigenstates is achieved by adapting methods from
the large deviations theory, in particular, the Varadhan lemma and
the Mogulskii theorem. Further details can be found in \cite{Franchini_Thesis,Franchini_URNS,Franchini_IRT},
where a full mathematical derivation is shown for the more general
HLS model (for example, the binary noise model is recovered in the
most trivial case of constant urn function).

\subsection{Lattice gas\protect\label{subsec:Lattice-gas}}

Notice that the set $\Omega\left(M\right)$ is equivalent to a self-avoiding
lattice gas of $E=N/2-M/2$ particles on a lattice of size $N$. Let
define the particle displacements 
\begin{equation}
X\left(\sigma_{V}\right):=\left\{ i\in V:\,\sigma_{i}=-1\right\} \subset V
\end{equation}
that in the magnetic representation would be the subset of $V$ where
the flickering state $\sigma_{V}^{*}$ is flipped with respect to
the master direction $1_{V}$ (that indicates a vector with all $1$
entries). Inside $\Omega\left(M\right)$ the size of $X$ is fixed,
and related to the magnetization by
\begin{equation}
|X\left(\sigma_{V}\right)|=\left(N-M\right)/2=:E,\ \ \forall\sigma_{V}\in\Omega\left(M\right).
\end{equation}
We interpret $E$ as the number of particles in our self--avoiding
gas. Then, we introduce the set of all possible displacements of $E=\left\lfloor \epsilon N\right\rfloor $
particles
\begin{equation}
\mathcal{V}\left(E\right):=\left\{ X\subset V:\,|X|=E\right\} ,\label{eq:LATTICEGAS}
\end{equation}
as for the magnetic representation before we use the notation
\begin{equation}
\mathcal{V}\left(\left\lfloor \epsilon N\right\rfloor \right)=:\mathcal{V}\left(\epsilon\right),\ \ \ \langle f\left(\sigma_{V}\right)\rangle_{\mathcal{V}\left(\left\lfloor \epsilon N\right\rfloor \right)}=:\langle f\left(\sigma_{V}\right)\rangle_{\epsilon}.
\end{equation}
that is in fact the exact image of $\Omega\left(m\right)$ if one
takes $\epsilon=\left(1-m\right)/2$. We represent the spin states
in terms of $X$ as follows: let $\sigma_{V}=\sigma_{V\setminus X}\cup\sigma_{X}$,
then for the flipped spins we have $\sigma_{X}=-1_{X}$, vector with
all negative entries, for the others $\sigma_{V\setminus X}=1_{V\setminus X}$
with all positive entries. The spin state $\sigma_{V}$ can be reconstructed
from the flipped vertices 
\begin{equation}
\sigma_{V}=1_{V\setminus X}\cup(-1_{X}),
\end{equation}
this representation in terms of particle displacements allows to easily
explore the overlap structure. Consider two configurations $X,Y\in\mathcal{V}\left(\epsilon\right)$,
corresponding to 
\begin{equation}
\sigma_{V}=1_{V\setminus X}\cup(-1_{X}),\ \ \tau_{V}=1_{V\setminus Y}\cup(-1_{Y}).
\end{equation}
Within $\mathcal{V}\left(\epsilon\right)$ the number of particles
is fixed $\left|X\right|=\left|Y\right|=\epsilon N$. Now, let $X\cap Y$
be the set of points of $V$ at which the two particle configurations
$X$ and $Y$ overlap, and define the \textbf{non-overlapping}\textbf{
components}
\begin{equation}
X':=X\setminus\left(X\cap Y\right),\ \ Y':=Y\setminus\left(X\cap Y\right),\label{eq:supergauge}
\end{equation}
that correspond to the non overlapping points of the particle displacements.
By definition, their intersection is void, ie $X'\cap Y'=\textrm{Ø}$,
moreover, the following equalities hold for of the union of $X$ and
$Y$ 
\begin{equation}
X\cup Y=X'\cup Y'\cup\left(X\cap Y\right),\ \ X'\cup Y'=\left(X\cup Y\right)\setminus\left(X\cap Y\right).\label{eq:supergauge1}
\end{equation}
It follows that $|X\cup Y|$ total fraction of $V$ occupied by the
particles is 
\begin{equation}
\left|X\cup Y\right|=\left|X\right|+\left|Y\right|-\left|X\cap Y\right|,
\end{equation}
while the total non-overlapping volume is
\begin{equation}
\left|X'\cup Y'\right|=\left|X\cup Y\right|-\left|X\cap Y\right|=\left|X\right|+\left|Y\right|-2\left|X\cap Y\right|.
\end{equation}
The overlap between the corresponding spin states can be expressed
as
\begin{multline}
\sigma_{V}\cdot\tau_{V}=\left[1_{V\setminus X}\cup(-1_{X})\right]\cdot\left[1_{V\setminus Y}\cup(-1_{Y})\right]=\\
=N-2|X'|-2|Y'|=N-2\left|X\right|-2\left|Y\right|+4\left|X\cap Y\right|,\label{eq:ffff-1}
\end{multline}
since we are considering magnetizations eigenstates with fixed eigenvalue,
the volumes of $\left|X\right|$ and $\left|Y\right|$ are also fixed
at $\epsilon N$, and the overlap of the spin states can be expressed
in terms of the overlap between the particle configurations:
\begin{equation}
\sigma_{V}\cdot\tau_{V}=\left(1-4\epsilon\right)N+4\left|X\cap Y\right|.
\end{equation}
The overlap size is $0\leq\left|X\cap Y\right|\leq\epsilon N$, but
it can be shown (e.g., see the next section) that for large $N$ the
overlap concentrates on $\epsilon^{2}N$, with fluctuations of order
$\sqrt{N}$ (it converges to a Gaussian), then from $m=1-2\epsilon$
and the Eq. (\ref{eq:ffff-1}) follows that the spin overlap concentrates
almost surely on the mean value $1-4\epsilon+4\epsilon^{2}=m^{2}$. 

\subsection{Entropy of the overlap\protect\label{subsec:Entropy-of-the}}

\begin{figure}
\includegraphics[scale=0.75]{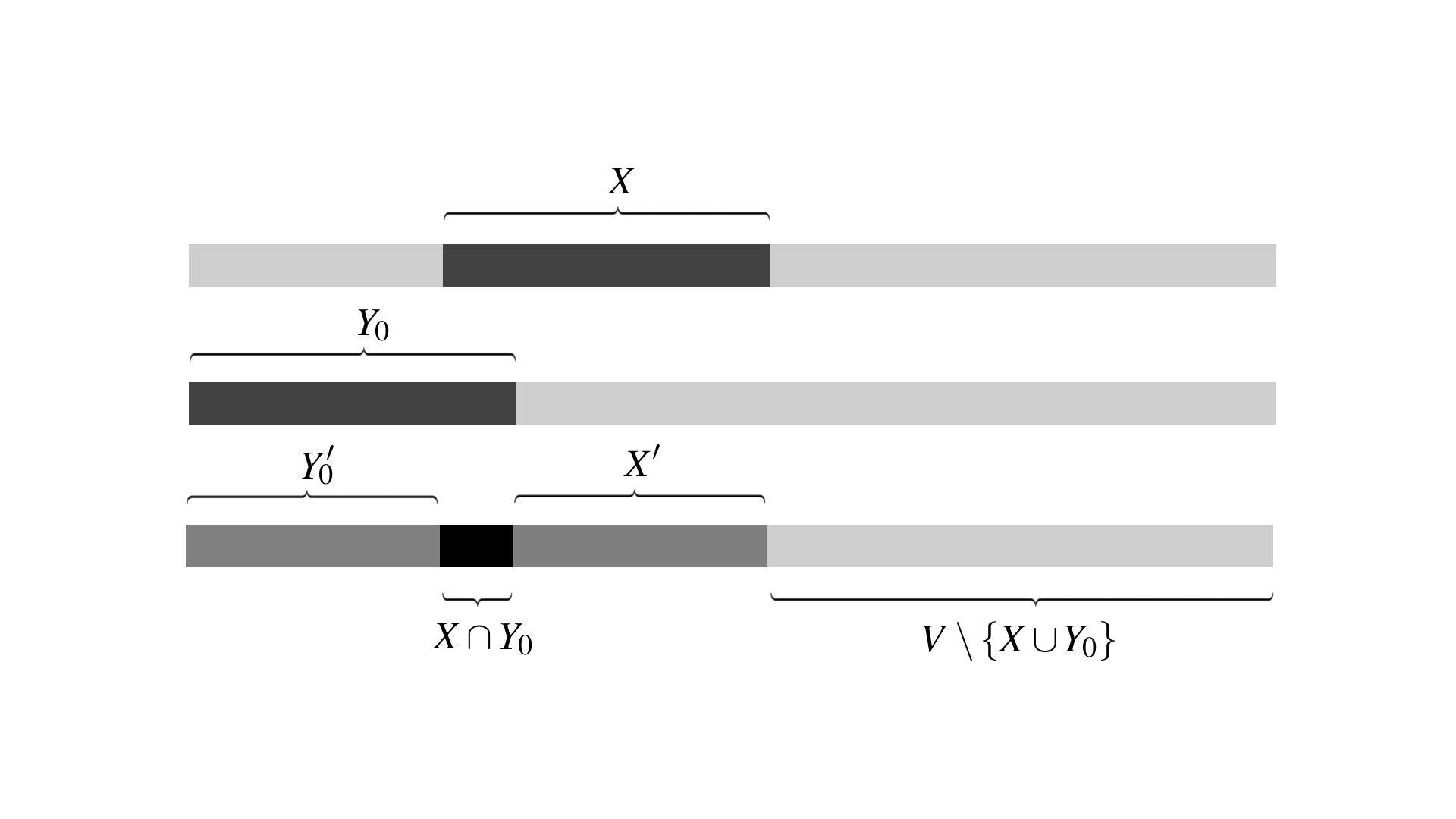}\caption{\protect\label{fig:A-configuration}Two configurations $X$ (first
row) and $Y_{0}$ (second row) extracted from $\mathcal{V}\left(\epsilon\right)$,
reordered in such way that both $X$ and $X\cap Y_{0}$ are compact
sets. The last row shows the partition into the disjoint non-overlapping
components $X'$, $Y'_{0}$ and the common component $X\cap Y_{0}$
projected on $V$ (last row).}
\end{figure}

We compute the probability that two configuration randomly extracted
from $\mathcal{V}\left(\epsilon\right)$ have an intersection of size
$\left\lfloor xN\right\rfloor $, with $x\in\left[0,\epsilon\right]$.
The limit entropy density (rate function) of such event is defined
as follows
\begin{equation}
\eta\left(x|\epsilon\right):=-\lim_{N\rightarrow\infty}\frac{1}{N}\log P\left(\,\left|X\cap Y\right|=\left\lfloor xN\right\rfloor \,|\,X,Y\in\mathcal{V}\left(\epsilon\right)\right)
\end{equation}
that gives the shape of the distribution for large number of spins
\begin{equation}
P\left(\,\left|X\cap Y\right|=\left\lfloor xN\right\rfloor \,|\,X,Y\in\mathcal{V}\left(\epsilon\right)\right)\sim\exp\left[-N\eta\left(x|\epsilon\right)\right].
\end{equation}
The first step is to notice that due to the uniformity of the distribution
of the flipped spins the intersection size does not depend on the
special realization of both states, then we can fix one of the two
states: let's fix $Y$ and call it `target' set, then
\begin{multline}
P\left(\,\left|X\cap Y\right|=\left\lfloor xN\right\rfloor \,|\,X,Y\in\mathcal{V}\left(\epsilon\right)\right)=\\
=P\left(\,\left|X\cap Y\right|=\left\lfloor xN\right\rfloor \,|\,X\in\mathcal{V}\left(\epsilon\right)\right),\ \forall Y\in\mathcal{V}\left(\epsilon\right).\label{eq:fffffff}
\end{multline}
Since only the size of the target set actually matters, to highlight
its internal components it will be convenient to chose a special configuration
of the target (see Figure \ref{fig:A-configuration})
\begin{equation}
Y_{0}:=\left\{ 1\leq k\leq\left\lfloor \epsilon N\right\rfloor \right\} 
\end{equation}
where the vertices of the flipped spins are placed at the beginning
of the set $V$ (ie, the labels $k\in V\setminus Y_{0}$ are all larger
than $\left\lfloor \epsilon N\right\rfloor $), formally holds that
\begin{equation}
P\left(\,\left|X\cap Y\right|=\left\lfloor xN\right\rfloor \,|\,X,Y\in\mathcal{V}\left(\epsilon\right)\right)=P\left(\,\left|X\cap Y_{0}\right|=\left\lfloor xN\right\rfloor \,|\,X\in\mathcal{V}\left(\epsilon\right)\right),
\end{equation}
The entropy density is given by the limit
\begin{equation}
\eta\left(x|\epsilon\right):=-\lim_{N\rightarrow\infty}\frac{1}{N}\log P\left(\,\left|X\cap Y_{0}\right|=\left\lfloor xN\right\rfloor \,|\,X\in\mathcal{V}\left(\epsilon\right)\right),
\end{equation}
and it can be computed in many ways by Varadhan Lemma, the Mogulskii
theorem and other large deviations techniques. 

\subsection{Urn methods\protect\label{subsec:Urn-methods}}

We can adapt methods from the urn process theory (see \cite{Franchini_URNS,Franchini_IRT})
to compute the shape of the overlap entropy density. The method consists
in defining a nested set sequence that start from the null set and
converges to $X$ in exactly $E$ steps,
\begin{equation}
X_{n}:=\bigcup_{n\leq E}\left\{ i_{n}\right\} ,
\end{equation}
that is a Markov chain with transition matrix
\begin{equation}
P\left(i_{n+1}=k\right)=\frac{\mathbb{I}\left(\,k\in V\setminus X_{n}\right)}{\left|V\setminus X_{n}\right|}.
\end{equation}
We indicate the overlap between $X_{n}$ and the target set $Y_{0}$
with 
\begin{equation}
R_{n}:=\left|X_{n}\cap Y_{0}\right|,
\end{equation}
the final conditions of the processes are fixed at $X_{E}=X$ and
$R_{E}=\left|X\cap Y_{0}\right|$ respectively, reached in $E=\left\lfloor \epsilon N\right\rfloor $
steps. It can be shown that the overlap between $X_{n}$ and the target
set follows a urn process \cite{Franchini_Thesis,Franchini_URNS,Franchini_IRT,Franchini_RW2,Jack}
in the step variable $n$
\begin{equation}
R_{n+1}=\left\{ \begin{array}{ccc}
R_{n}+1 &  & \pi_{n}\left(R_{n}\right)\\
R_{n} &  & 1-\pi_{n}\left(R_{n}\right)
\end{array}\right.
\end{equation}
the urn function $\pi_{n}$ at step $n$ is the ratio between the
number of vertices in $Y_{0}$ that have not been occupied in the
preceding $n$ extractions (that are $E-R_{n}$), and the number of
vertices of $V$ that have not been occupied (ie, $N-n$),
\begin{equation}
\pi_{n}\left(R_{n}\right):=\frac{E-R_{n}}{N-n}.
\end{equation}
adapting large-deviation methods \cite{Franchini_Thesis,Franchini_URNS,Franchini_IRT,Franchini_RW2,Jack}
from generalized urn models is possible to show that the distribution
of the overlap is approximately Gaussian. It is also possible to compute
the parameters by solving the difference equation
\begin{equation}
\mathbb{E}\left(R_{n+1}\right)=\mathbb{E}\left[\left(R_{n}+1\right)\pi_{n}\left(R_{n}\right)\right]+\mathbb{E}\left\{ R_{n}\left[1-\pi_{n}\left(R_{n}\right)\right]\right\} 
\end{equation}
where $\mathbb{E}\left(\,\cdot\,\right)$ indicates the average respect
to the urn process. Substituting the expression of the urn function
we find
\begin{equation}
\mathbb{E}\left(R_{n+1}\right)-\mathbb{E}\left(R_{n}\right)=-\frac{1}{N-n}\mathbb{E}\left(R_{n}\right)+\frac{E}{N-n},
\end{equation}
solving with null initial condition brings to the linear average solution
$\mathbb{E}\left(R_{n}\right)=nE/N$ from which follows that the average
overlap converges to 
\begin{equation}
\lim_{N\rightarrow\infty}\frac{\langle\left|X\cap Y_{0}\right|\rangle_{\epsilon}}{N}=\lim_{N\rightarrow\infty}\frac{\mathbb{E}\left(R_{E}\right)}{N}=\epsilon^{2}.
\end{equation}
We can show that the fluctuations are small: consider 
\begin{equation}
\mathbb{E}\left(R_{n+1}^{2}\right)=\mathbb{E}[\left(R_{n}+1\right)^{2}\pi_{n}\left(R_{n}\right)]+\mathbb{E}\left\{ R_{n}^{2}\left[1-\pi_{n}\left(R_{n}\right)\right]\right\} ,
\end{equation}
substituting the urn function and the formula for the linear we find
\begin{equation}
\mathbb{E}\left(R_{n+1}^{2}\right)-\mathbb{E}\left(R_{n}^{2}\right)=-\frac{2}{N-n}\,\mathbb{E}\left(R_{n}^{2}\right)+\frac{E}{N-n}\left[1+n\left(\frac{2E-1}{N}\right)\right],
\end{equation}
solving again for null initial condition gives another linear solution
\begin{equation}
\mathbb{E}\left(R_{n}^{2}\right)=\frac{nE\left[E\left(n-1\right)-n+N\right]}{N\left(N-1\right)}
\end{equation}
Let now compute the variance of $R_{n}$: the variance is defined
by
\begin{equation}
\mathbb{E}\left(R_{n}^{2}\right)-\mathbb{E}\left(R_{n}\right)^{2}=\frac{nE\left[E\left(n-1\right)-n+N\right]}{N\left(N-1\right)}-\frac{n^{2}E^{2}}{N^{2}}
\end{equation}
and after some algebra it can be shown that for the variance holds
\begin{equation}
\lim_{N\rightarrow\infty}\frac{\langle\left|X\cap Y_{0}\right|^{2}\rangle_{\epsilon}-\langle\left|X\cap Y_{0}\right|\rangle_{\epsilon}^{2}}{N}=\lim_{N\rightarrow\infty}\frac{\mathbb{E}\left(R_{E}^{2}\right)-\mathbb{E}\left(R_{E}\right)^{2}}{N}=\epsilon^{2}\left(1-\epsilon\right)^{2}.
\end{equation}
The entropy density $\eta$ can thus be expanded at second order in
the variable
\begin{equation}
\Lambda\left(x|\epsilon\right):=\frac{x-\epsilon^{2}}{\epsilon\left(1-\epsilon\right)},\label{eq:renorma}
\end{equation}
in the limit of large $N$ it can be shown that 
\begin{equation}
\eta\left(x|\epsilon\right)=\frac{1}{2}\,\Lambda\left(x|\epsilon\right)^{2}+O[\,\Lambda\left(x|\epsilon\right)^{3}],
\end{equation}
and since $m=1-2\epsilon$, according to Eq. (\ref{eq:ffff-1}), the
corresponding spin overlap concentrates almost surely on the average
value, ie., $1-4\epsilon+4\epsilon^{2}=m^{2}$. Notice that the spin
overlap concentrates on the same value of the correlation matrix $\langle\sigma_{i}\sigma_{j}\rangle_{m}=m^{2}$.
This means that the kernel of the magnetization eigenstates commutes
in distribution, ie., that the correlation matrix converges to the
overlap matrix, see in Section 2 of \cite{Franchini2023} for further
details on kernel commutation and its implications.

\section{Binary noise model\protect\label{sec:Binary-noise-model}}

Let now consider the simplest situation where the field has two states
only (binary noise model), the absolute value is a delta function
centered on one, that is $p\left(x\right)=\delta\left(x-1\right)$.
We study the Hamiltonian $\sigma_{V}\cdot\omega_{V}$, scalar product
between $\sigma_{V}$ and the input $\omega_{V}$, 
\begin{equation}
H\left(\sigma_{V}|\omega_{V}\right)=\sum_{i\in V}\sigma_{i}\omega_{i}.
\end{equation}
Since $|\omega_{i}|=1$ the canonical analysis here is very simple:
notice that due to parity of the $\cosh$ function the partition function
does not depend on the input state $\omega_{V}$, 
\begin{equation}
Z\left(\beta\right)=\sum_{\sigma_{V}\in\Omega^{V}}\exp\,\left(-\beta\sigma_{V}\cdot\omega_{V}\right)=\left[2\cosh\left(\beta\right)\right]^{N},
\end{equation}
the free energy per spin and the ground state energy are 
\begin{equation}
-\beta\zeta\left(\beta\right)=\log2\cosh\left(\beta\right),\ \ \ \psi=\lim_{\beta\rightarrow\infty}\tanh\left(\beta\right)=1.
\end{equation}

\subsection{Free energy phases\protect\label{subsec:Free-energy-phases}}

In the low temperature limit the free energy is 
\begin{equation}
-\beta\zeta\left(\beta\right)=\beta+\exp\left(-2\beta\right)+O\left[\exp\,(-4\beta)\right],
\end{equation}
then, the free energy per spin converges to the ground state energy
$\zeta\left(\infty\right)$ exponentially fast in $\beta$. Moreover,
we find at high temperature the free energy converges to the replica
symmetric (RS) free energy of the spin glass theory: Taylor expansion
of the $\log\cosh$ function for small $\beta$ gives 
\begin{equation}
-\beta\zeta\left(\beta\right)=\log2+\frac{\beta^{2}}{2}+O\left(\beta^{4}\right).
\end{equation}
It can be shown that in the zero temperature limit the Gibbs measure
can be approximated by a random energy model: this will be discussed
later. 

\subsection{Flickering states and thermal average\protect\label{subsec:Staggering-states-and}}

Let study the formula for the average:
\begin{equation}
\langle f\left(\sigma_{V}\right)\rangle_{\xi}=\frac{1}{Z\left(\beta\right)}\sum_{\sigma_{V}\in\Omega^{V}}f\left(\sigma_{V}\right)\,\exp\left(-\beta\sigma_{V}\cdot\omega_{V}\right).
\end{equation}
Given the independence of the partition function from $\omega_{V}$
it will be convenient to introduce some notation. Define the flickering
state $\sigma_{V}^{*}:=\sigma_{V}\circ\omega_{V}$ such that the resulting
vector has the following components $\sigma_{i}^{*}:=\sigma_{i}\,\omega_{i}\in\Omega$.
Notice that since $\omega_{i}^{2}=1$ a further multiplication of
$\sigma_{V}^{*}$ by $\omega_{V}$ gives back the original vector
$\sigma_{V}$, ie., $\sigma_{V}^{*}\circ\omega_{V}=\sigma_{V}$. Then
we introduce the flickering function
\begin{equation}
f^{*}\left(\sigma_{V}\right):=f\left(\sigma_{V}\circ\omega_{V}\right),\ \ \ f^{*}\left(\sigma_{V}^{*}\right)=f\left(\sigma_{V}^{*}\circ\omega_{V}\right)=f\left(\sigma_{V}\right).
\end{equation}
Finally, we consider the scalar product (overlap) of $\sigma_{V}$
with the input state $\omega_{V}$, that is equivalent to the total
magnetization of $\sigma_{V}^{*}$, 
\begin{equation}
\sigma_{V}\cdot\omega_{V}=\left(\sigma_{V}\circ\omega_{V}\right)\cdot1_{V}=\sigma_{V}^{*}\cdot1_{V}=:M\left(\sigma_{V}^{*}\right).
\end{equation}
Putting together, the sum of $f$ weighted with the Gibbs weights
satisfies the following chain of equivalences 
\begin{multline}
\sum_{\sigma_{V}\in\Omega^{V}}f\left(\sigma_{V}\right)\exp\left(-\beta\sigma_{V}\cdot\omega_{V}\right)=\\
=\sum_{\sigma_{V}\in\Omega^{V}}f\left(\sigma_{V}^{*}\circ\omega_{V}\right)\exp\left[-\beta M\left(\sigma_{V}^{*}\right)\right]=\sum_{\sigma_{V}\in\Omega^{V}}f^{*}\left(\sigma_{V}^{*}\right)\exp\left[-\beta M\left(\sigma_{V}^{*}\right)\right]=\\
=\sum_{\sigma_{V}\in\Omega^{V}}f^{*}\left(\sigma_{V}\right)\exp\left[-\beta M\left(\sigma_{V}\right)\right],\label{eq:dfdfdfdfd}
\end{multline}
where in the last step we used that $\sigma_{V}$ and $\sigma_{V}^{*}$
are in a bijective relation, this implies that we can change the sum
index to $\sigma_{V}$ as the dependence on the input state affects
only $f^{*}$. Then, the formula for the average is as follows:
\begin{multline}
\langle f\left(\sigma_{V}\right)\rangle_{\xi}=\frac{1}{Z\left(\beta\right)}\sum_{\sigma_{V}\in\Omega^{V}}\exp\left[-\beta M\left(\sigma_{V}^{*}\right)\right]f^{*}\left(\sigma_{V}\right)=\\
=\frac{1}{Z\left(\beta\right)}\sum_{M}\exp\left(-\beta M\right)\sum_{\sigma_{V}\in\Omega\left(M\right)}f^{*}\left(\sigma_{V}\right)=\\
=\frac{1}{Z\left(\beta\right)}\sum_{M}|\,\Omega\left(M\right)|\exp\left(-\beta M\right)\langle f^{*}\left(\sigma_{V}\right)\rangle_{\Omega\left(M\right)}.\label{eq:ddsdddsd}
\end{multline}

\subsection{Average in thermodynamic limit\protect\label{subsec:Average-in-thermodynamic}}

Assuming that $f$ exists in the thermodynamic limit $N\rightarrow\infty$,
we can write also a continuous representation. From \cite{Dembo_Zeitouni,Franchini_Thesis,Franchini_URNS,Franchini_IRT,Franchini_RW2,Jack}
it can be shown that
\begin{equation}
\lim_{N\rightarrow\infty}\langle f\left(\sigma_{V}\right)\rangle_{\xi}=\lim_{N\rightarrow\infty}\frac{\int_{-1}^{1}dm\,\exp\left\{ N\left[\phi\left(m\right)-\beta m\right]\right\} \langle f^{*}\left(\sigma_{V}\right)\rangle_{m}}{\int_{-1}^{1}dm\,\exp\left\{ N\left[\phi\left(m\right)-\beta m\right]\right\} }
\end{equation}
and it can be also shown that the probability mass concentrates on
the value $m_{0}\left(\beta\right)$ that maximize $\beta m+\phi\left(m\right)$.
Putting together 
\begin{equation}
\lim_{N\rightarrow\infty}\langle f\left(\sigma_{V}\right)\rangle_{\xi}=\lim_{N\rightarrow\infty}\langle f^{*}\left(\sigma_{V}\right)\rangle_{m_{0}\left(\beta\right)}
\end{equation}
and after some manipulations one can prove that $m_{0}\left(\beta\right)=\tanh\left(\beta\right)$.
Then, it is possible to compute the average in terms of the eigenstates
of magnetization and their effects on the flickering function $f^{*}$.

\section{Relation with the Random Energy Model\protect\label{sec:Relation-with-the}}

It can be shown that at low temperature the Gibbs measure converges
in distribution to a Random Energy Model (REM) of the Derrida type
\cite{Franchini2023}. Define 
\begin{equation}
\psi=\frac{1}{N}\sum_{i\in V}x_{i},\ \ \ \varphi_{i}:=x_{i}-\psi,
\end{equation}
where $\psi$ is the ground state energy and $\varphi_{i}$ describes
the field fluctuations. The Hamiltonian can be rewritten once again
as follows
\begin{equation}
\sigma_{V}\cdot h_{V}=\psi\,M(\sigma_{V}^{*})+\sigma_{V}^{*}\cdot\varphi_{V}.
\end{equation}
As in previous section, we recall the special notation for the composition
between the master direction and the test function, we called it flickering
function
\begin{equation}
f^{*}\left(\sigma_{V}\right)=f\left(\sigma_{V}\circ\omega_{V}\right),
\end{equation}
and notice that it does not depend on the external field $x_{V}$.
Then, the average is rewritten in terms of the flickering variables
only
\begin{equation}
\langle f\left(\sigma_{V}\right)\rangle_{\xi}=\sum_{\sigma_{V}\in\Omega^{V}}f^{*}\left(\sigma_{V}\right)\,\exp\left[-\beta\psi\,M(\sigma_{V})-\beta\sigma_{V}\cdot\varphi_{V}+N\beta\zeta\left(\beta x_{V}\right)\right]
\end{equation}
so that the dependence on $\omega_{V}$ is all inside the flickering
function $f^{*}$ and both the free energy density and the Gibbs measure
depend only on the rectified field $x_{V}$. 

\subsection{Field fluctuations revisited\protect\label{subsec:Field-fluctuations-1}}

By the lattice gas representation described before, the following
holds:
\begin{equation}
\frac{\sigma_{V}\cdot\varphi_{V}}{2}=-1_{X(\sigma_{V})}\cdot\,\varphi_{X(\sigma_{V})},
\end{equation}
in fact, consider the chain of identities
\begin{equation}
\sum_{i\in V}\sigma_{i}\varphi_{i}=\sum_{i\in V}\varphi_{i}-\sum_{i\in V}\left(1-\sigma_{i}\right)\varphi_{i}=\sum_{i\in V}\varphi_{i}-\sum_{i\in X(\sigma_{V})}2\varphi_{i},
\end{equation}
by definition we have that the first sum is zero, 
\begin{equation}
\sum_{i\in V}\varphi_{i}=\sum_{i\in V}x_{i}-\sum_{i\in V}\psi=0.
\end{equation}
Let now introduce a notation for the variance inside $\varphi_{V}$,
that we denote by $\delta^{2}$, and the variance over the vertex
set $\gamma^{2}$
\begin{equation}
\delta^{2}=\frac{1}{N}\sum_{i\in V}\varphi_{i}^{2},\ \ \ \gamma^{2}=\frac{1}{N}\sum_{i\in V}x_{i}^{2}
\end{equation}
this quantity is related to ground state and variance of $x_{i}^{2}$
by the relation $\delta^{2}=\gamma^{2}-\psi^{2}$ where $\gamma$
is the average variance over the vertex set. 

\subsection{The $J-$field\protect\label{subsec:The-field}}

We can now introduce a fundamental variable, that we call $J-$field
\begin{equation}
J_{X(\sigma_{V})}:=\left\{ J_{i}:=\varphi_{i}/\delta:\,i\in X(\sigma_{V})\right\} ,
\end{equation}
from which we define the normalized field amplitude
\begin{equation}
J(\sigma_{V}):=\frac{1_{X(\sigma_{V})}\cdot\,J_{X(\sigma_{V})}}{\sqrt{|X(\sigma_{V})|}}.
\end{equation}
This variable converges to a Gaussian with zero mean and unitary variance
in the thermodynamic limit, moreover, given two states $\sigma{}_{V}$
and $\tau{}_{V}$ independently extracted from $\Omega\left(m\right)$
the average overlap converges to $(1-m)/2$. From previous considerations
the Hamiltonian can be rewritten as follows
\begin{equation}
\sigma_{V}\cdot h_{V}=\psi\,M(\sigma_{V}^{*})+\,J(\sigma_{V}^{*})\sqrt{K\left[\,M(\sigma_{V}^{*})\right]},
\end{equation}
where we introduced a notation for the normalization of the $J-$amplitude
\begin{equation}
K\left[\,M(\sigma_{V})\right]:=2\delta^{2}N\left[\,1-M(\sigma_{V})/N\,\right].
\end{equation}
The formula for the average is rewritten in terms of the new variables
\begin{multline}
\langle f\left(\sigma_{V}\right)\rangle_{\xi}=\\
=\frac{\sum_{\,M}|\Omega\left(M\right)|\exp\left(-\beta\psi\,M\right)\langle f^{*}\left(\sigma_{V}\right)\,\exp\,[-\beta J(\sigma_{V})\sqrt{K(M)}]\rangle_{\Omega\left(M\right)}}{\sum_{\,M}|\Omega\left(M\right)|\exp\left(-\beta\psi\,M\right)\langle\,\exp\,[-\beta J(\sigma_{V})\sqrt{K(M)}]\rangle_{\Omega\left(M\right)}}.\label{eq:fffffddddd}
\end{multline}
Now, let take the thermodynamic limit: it can be shown by simple saddle
point methods \cite{Franchini_URNS,Franchini_IRT} that the average
admit the following integral representation
\begin{multline}
\lim_{N\rightarrow\infty}\langle f\left(\sigma_{V}\right)\rangle_{\xi}=\\
=\lim_{N\rightarrow\infty}\frac{\int_{-1}^{1}dm\,\exp\left\{ N[\phi\left(m\right)-\beta\psi m]\right\} \,\langle f^{*}\left(\sigma_{V}\right)\,\exp\,[-\beta J(\sigma_{V})\sqrt{K\left(m\right)}]\rangle_{m}}{\int_{-1}^{1}dm\,\exp\left\{ N[\phi\left(m\right)-\beta\psi m]\right\} \,\langle\exp\,[-\beta J(\sigma_{V})\sqrt{K\left(m\right)}]\rangle_{m}}\label{eq:wergfaga-1}
\end{multline}
introducing the auxiliary function
\begin{equation}
K\left(m\right):=2\delta^{2}\left(1-m\right)N,
\end{equation}
we arrive to the final form for our average formula, that is
\begin{multline}
\lim_{N\rightarrow\infty}\langle f\left(\sigma_{V}\right)\rangle_{\xi}=\lim_{N\rightarrow\infty}\frac{\langle f^{*}\left(\sigma_{V}\right)\,\exp\,[-\beta J(\sigma_{V})\sqrt{K\left(m_{0}\right)}]\rangle_{m_{0}}}{\langle\,\exp\,[-\beta J(\sigma_{V})\sqrt{K\left(m_{0}\right)}]\rangle_{m_{0}}}=\\
=\lim_{N\rightarrow\infty}\frac{\sum_{\,\sigma_{V}\in\Omega\left(m_{0}\right)}f^{*}\left(\sigma_{V}\right)\,\exp\,[-\beta J(\sigma_{V})\sqrt{K\left(m_{0}\right)}]}{\sum_{\,\sigma_{V}\in\Omega\left(m_{0}\right)}\,\exp\,[-\beta J(\sigma_{V})\sqrt{K\left(m_{0}\right)}]}.\label{eq:averagefinalformula}
\end{multline}
In \cite{Franchini2023} is shown that in the low temperature limit
the Gaussian amplitude $J$ converges in the bulk to a random energy
model of the Derrida type \cite{Kurkova1} (ie., with Gaussian energies).
This is done by noticing that when the temperature goes to zero the
state aligns toward the direction of the ground state almost everywhere,
and only a small fraction of spins get flipped in the opposite direction.
Since the flipped spins are sparse, any two independent configurations
will most probably have a negligible number of common flips. The number
of this common flips (see Figure \ref{fig:A-configuration}) converges
to zero faster than the size of the whole flipped set when the temperature
is lowered to near zero (ie., net of quadratic terms), and can be
therefore ignored in that limit: see Section 5 of \cite{Franchini2023}
for further details. The crucial fact is in that the field $J_{i}$
is sampled independently for each vertex $i$, then for any two disjoint
subsets of $V$ the corresponding $J-$fields are independent like
in a REM. The argument works also for multiple replicas if temperature
is low enough.

\subsection{REM at all temperatures\protect\label{subsec:The-field-1}}

In this last sub-section we show how is possible to correct the formulas
of \cite{Franchini2023} in order to make it valid also at higher
temperatures. Let consider two subsets of $X,Y\subset V$ of same
size $E$ and their non-overlapping components $X'$ and $Y'$ as
defined in Eq. (\ref{eq:supergauge}) of Section 4. Now notice that
the following holds:
\begin{equation}
X=X'\cup\{X\cap Y\},\ \ \ Y=Y'\cup\{X\cap Y\}.
\end{equation}
The the REM contribution comes only from the non-overlapping components,
then we would like to get rid of the overlapping component (ie, the
energy of the spins placed on the vertices in $X\cap Y$ ) and write
everything in terms of the sets $X'$ and $Y'$ . This is made possible
by considering the difference between the corresponding $J-$fields
\begin{multline}
1_{X}\cdot J_{X}-1_{Y}\cdot J_{Y}=\left(1_{X'}\cdot J_{X'}+1_{X\cap Y}\cdot J_{X\cap Y}\right)-\left(1_{Y'}\cdot J_{Y'}+1_{X\cap Y}\cdot J_{X\cap Y}\right)=\\
=1_{X'}\cdot J_{X'}-1_{Y'}\cdot J_{Y'}\label{eq:renormalization}
\end{multline}
Therefore, let consider two independent replicas $\sigma_{V}$ and
$\tau_{V}$ and let indicate with $X(\sigma_{V})$ and $Y(\tau_{V})$
the associated flipped components. Let introduce the auxiliary $\Delta-$field,
that is the difference between the $J-$fields of the two replicas
\begin{equation}
\Delta\left(\sigma_{V}|\tau_{V}\right):=J\left(\sigma_{V}\right)-J\left(\tau_{V}\right),
\end{equation}
by multiplying both numerator and denominator of the average formula
in Eq. (\ref{eq:averagefinalformula}) by the proper $\tau_{V}-$dependent
amplitude: we find
\begin{multline}
\frac{\sum_{\sigma_{V}\in\Omega\left(m_{0}\right)}f^{*}\left(\sigma_{V}\right)\exp\,[-\beta J\left(\sigma_{V}\right)\sqrt{K\left(m_{0}\right)}]}{\sum_{\sigma_{V}\in\Omega\left(m_{0}\right)}\exp\,[-\beta J\left(\sigma_{V}\right)\sqrt{K\left(m_{0}\right)}]}=\\
=\frac{\sum_{\sigma_{V}\in\Omega\left(m_{0}\right)}f^{*}\left(\sigma_{V}\right)\exp\,\{-\beta\left[J\left(\sigma_{V}\right)-J\left(\tau_{V}\right)\right]\sqrt{K\left(m_{0}\right)}\}}{\sum_{\sigma_{V}\in\Omega\left(m_{0}\right)}\exp\,\{-\beta[J\left(\sigma_{V}\right)-J\left(\tau_{V}\right)]\sqrt{K\left(m_{0}\right)}\}}=\\
=\frac{\sum_{\sigma_{V}\in\Omega\left(m_{0}\right)}f^{*}\left(\sigma_{V}\right)\exp\,[-\beta\Delta\left(\sigma_{V}|\tau_{V}\right)\sqrt{K\left(m_{0}\right)}]}{\sum_{\sigma_{V}\in\Omega\left(m_{0}\right)}\exp\,[-\beta\Delta\left(\sigma_{V}|\tau_{V}\right)\sqrt{K\left(m_{0}\right)}]}\label{eq:renormalization-1}
\end{multline}
from previous considerations and Eq. (\ref{eq:renormalization}) is
easy to verify that the overlapping component cancels out and
\begin{multline}
\Delta\left(\sigma_{V}|\tau_{V}\right)=\frac{1}{\sqrt{E}}[1_{X\left(\sigma_{V}\right)}\cdot J_{X\left(\sigma_{V}\right)}-1_{Y\left(\tau_{V}\right)}\cdot J_{Y\left(\tau_{V}\right)}]=\\
=\frac{1}{\sqrt{E}}[1_{X'\left(\sigma_{V}\right)}\cdot J_{X'\left(\sigma_{V}\right)}-1_{Y'\left(\tau_{V}\right)}\cdot J_{Y'\left(\tau_{V}\right)}]=\\
=\sqrt{\frac{E'}{E}}[J'\left(\sigma_{V}\right)-J'\left(\tau_{V}\right)]=\sqrt{\frac{E'}{E}}\,\Delta'\left(\sigma_{V}|\tau_{V}\right),\label{eq:renormalization-1-1}
\end{multline}
Now, since $E'/E$ converges to $1-\epsilon_{0}$ in the thermodynamic
limit we have
\begin{equation}
J\left(\sigma_{V}\right)=\sqrt{1-\epsilon_{0}}\ J'\left(\sigma_{V}\right)
\end{equation}
Most important: notice that the $J'-$amplitude is distributed like
a REM by construction since we obtained it by removing the ``non-REM''
component from $J$. Then, let define one last auxiliary function
\begin{multline}
K'\left(m_{0}\right):=\left(1-\epsilon_{0}\right)K\left(m_{0}\right)=\left[1-\left(1-m_{0}\right)/2\,\right][2\delta^{2}\left(1-m_{0}\right)N]=\\
=\delta^{2}\left(1+m_{0}\right)\left(1-m_{0}\right)N=\delta^{2}\left(1-m_{0}^{2}\right)N\label{eq:renormalization-2}
\end{multline}
and put everything together, the average formula can be transformed
into
\begin{multline}
\lim_{N\rightarrow\infty}\langle f\left(\sigma_{V}\right)\rangle_{\xi}=\frac{\sum_{\sigma_{V}\in\Omega\left(m_{0}\right)}f^{*}\left(\sigma_{V}\right)\exp\,[-\beta J'\left(\sigma_{V}\right)\sqrt{K'\left(m_{0}\right)}]}{\sum_{\sigma_{V}\in\Omega\left(m_{0}\right)}\exp\,[-\beta J'\left(\sigma_{V}\right)\sqrt{K'\left(m_{0}\right)}]}\\
=\lim_{N\rightarrow\infty}\frac{\langle f^{*}\left(\sigma_{V}\right)\,\exp\,[-\beta J'(\sigma_{V})\sqrt{K'\left(m_{0}\right)}]\rangle_{m_{0}}}{\langle\exp\,[-\beta J'(\sigma_{V})\sqrt{K'\left(m_{0}\right)}\rangle_{m_{0}}}.\label{eq:averagefinalformula-1}
\end{multline}
We can immediately verify that after this change of variable the average
is done respect to a field with zero overlap matrix at any temperature.

\subsection{REM-PPP average \protect\label{subsec:REM-PPP-average}}

We can integrate the REM variable $J'$  by applying the well known
REM-PPP average formula \cite{Franchini2021,Franchini2023,ASS,Kurkova1}.
The final result is the relation in Eq. (\ref{eq:REM-PPP}) 
\begin{equation}
\lim_{N\rightarrow\infty}\langle f\left(\sigma_{V}\right)\rangle_{\xi}=\lim_{N\rightarrow\infty}\langle f^{*}\left(\sigma_{V}\right)^{\lambda}\rangle_{m_{0}}^{1/\lambda}
\end{equation}
with $\lambda$ depending on $\beta$, $\psi$ and $\delta$. Notice
that the REM-PPP average formula interpolates between arithmetic and
geometric average, in fact, 
\begin{equation}
\lim_{\lambda\rightarrow1}\langle f\left(\sigma_{V}\right)^{\lambda}\rangle_{m_{0}}^{1/\lambda}=\langle f\left(\sigma_{V}\right)\rangle_{m_{0}}=\frac{1}{|\Omega\left(m_{0}\right)|}\sum_{\sigma_{V}\in\Omega\left(m_{0}\right)}f\left(\sigma_{V}\right),
\end{equation}
and with little more work it is possible to show that
\begin{multline}
\lim_{\lambda\rightarrow0}\langle f\left(\sigma_{V}\right)^{\lambda}\rangle_{m_{0}}^{1/\lambda}=\lim_{\lambda\rightarrow0}\left[\frac{1}{|\Omega\left(m_{0}\right)|}\sum_{\sigma_{V}\in\Omega\left(m_{0}\right)}f\left(\sigma_{V}\right)^{\lambda}\right]^{1/\lambda}=\\
\ \ \ \ \ \ \,=\lim_{\lambda\rightarrow0}\left[\frac{1}{|\Omega\left(m_{0}\right)|}\sum_{\sigma_{V}\in\Omega\left(m_{0}\right)}\exp\,\lambda\log f\left(\sigma_{V}\right)\right]^{1/\lambda}=\\
\ \ \,\,=\lim_{\lambda\rightarrow0}\left[1+\frac{\lambda}{|\Omega\left(m_{0}\right)|}\sum_{\sigma_{V}\in\Omega\left(m_{0}\right)}\log f\left(\sigma_{V}\right)\right]^{1/\lambda}=\\
\,=\lim_{\lambda\rightarrow0}\left[1+\lambda\log\prod_{\sigma_{V}\in\Omega\left(m_{0}\right)}f\left(\sigma_{V}\right)^{\frac{1}{|\Omega\left(m_{0}\right)|}}\right]^{1/\lambda}=\\
\,=\lim_{\lambda\rightarrow0}\left[\exp\,\lambda\log\prod_{\sigma_{V}\in\Omega\left(m_{0}\right)}f\left(\sigma_{V}\right)^{\frac{1}{|\Omega\left(m_{0}\right)|}}\right]^{1/\lambda}=\\
=\prod_{\sigma_{V}\in\Omega\left(m_{0}\right)}f\left(\sigma_{V}\right)^{\frac{1}{|\Omega\left(m_{0}\right)|}},\label{eq:zdfgvzdedf}
\end{multline}
that is the geometric average. These formulas allows to computate
of the average with respect to the thermal fluctuations at any temperature
in the Gaussian case. Although notice: the dependence of $f^{*}$
from the ground state is still present. See Lemma 13 in Section 5
of Ref. \cite{Franchini2023} for further details on how to actually
compute $\lambda$ in terms of $\beta$, $\psi$ and $\delta$ in
the Gaussian case (or in the low temperature limit). 

Anyway, notice that the $J'$ field is only approximately Gaussian,
and its rate function \cite{Dembo_Zeitouni} could be different from
a quadratic form when the field fluctuations are large. The reason
why at low temperatures one can actually consider the bulk (which
makes the arguments relatively elementary) is in that the contributions
from spins with a near zero external field is only quadratic in temperature,
as shown in Section \ref{sec:Thermodynamic-limit} for the Gaussian
and uniform cases. 

This remarkable fact guarantees that the approximate Gaussianity of
$J'$ works up to the linear order in temperature, and then $J'$
is properly approximated by a REM of the Derrida type (i.e., with
Gaussian energies) in that limit. More general forms of REM should
be considered if we are interested in extending the computation of
$\lambda$ presented in Section 5 of \cite{Franchini2023} to non--Gaussian
fields and in the full temperature range, like those studied by N.
K. Jana in \cite{Kumar=000020Jana} that admit random energies with
arbitrary large deviation profile. This will be addressed elsewhere. 

\subsection{Evaluation of $m_{0}$\protect\label{subsec:Evaluation-of-emm}}

\noindent We will now evaluate the function $m_{0}$: respect to what
already computed in Section \ref{sec:Binary-noise-model}, the contribution
due to the random energy term in Eq. (\ref{eq:wergfaga-1}) is of
the same magnitude of the non--random term, and should be considered
when computing the shape of the function $m_{0}$. Therefore, the
shape of $m_{0}$ is not the hyperbolic tangent, but an slightly different
function. We restart from Eq. (\ref{eq:wergfaga-1}): 
\begin{multline}
\lim_{N\rightarrow\infty}\langle f\left(\sigma_{V}\right)\rangle_{\xi}=\\
=\lim_{N\rightarrow\infty}\frac{\int_{-1}^{1}dm\,\exp\left\{ N\left[\phi\left(m\right)-\beta\psi m\right]\right\} \langle f^{*}\left(\sigma_{V}\right)\exp\,[-\beta J'\left(\sigma_{V}\right)\sqrt{K'\left(m\right)}]\rangle_{m}}{\int_{-1}^{1}dm\,\exp\left\{ N\left[\phi\left(m\right)-\beta\psi m\right]\right\} \langle\exp\,[-\beta J'\left(\sigma_{V}\right)\sqrt{K'\left(m\right)}]\rangle_{m}}\label{eq:afd}
\end{multline}
we considered directly the re--normalized amplitude $K'$ of Eq.s
(\ref{eq:fianeq}) and (\ref{eq:renormalization-2}) since it eventually
converges to $K$ in the low temperature limit. The field $J'$ is
a random energy and $\phi$ is the binary entropy of Eq. (\ref{eq:eigenstatesrate}). 

To compute $m_{0}$ we study the normalization of Eq. (\ref{eq:wergfaga-1}):
the cardinality (complexity) of the magnetization ensemble with eigenvalue
$m$ is asymptotically
\begin{equation}
|\,\Omega\left(m\right)|\sim\exp\left[\,N\phi\left(m\right)\right].
\end{equation}
The definition of ensemble average is:
\begin{equation}
\langle\exp\,[-\beta J'\left(\sigma_{V}\right)\sqrt{K'\left(m\right)}]\rangle_{m}=\frac{1}{|\,\Omega\left(m\right)|}\sum_{\sigma_{V}\in\Omega\left(m\right)}\exp\,[-\beta J'\left(\sigma_{V}\right)\sqrt{K'\left(m\right)}].
\end{equation}
By substituting in the normalization of Eq. (\ref{eq:afd}) the two
$\phi$ cancel out:
\begin{multline}
\int_{-1}^{1}dm\,\exp\left\{ N\left[\phi\left(m\right)-\beta\psi m\right]\right\} \langle\exp\,[-\beta J'\left(\sigma_{V}\right)\sqrt{K'\left(m\right)}]\rangle_{m}=\\
=\int_{-1}^{1}dm\,\exp\left(-N\beta\psi m\right)\sum_{\sigma_{V}\in\Omega\left(m\right)}\exp\,[-\beta J'\left(\sigma_{V}\right)\sqrt{K'\left(m\right)}].\label{eq:hgjg}
\end{multline}
We can now regroup the constants and use the scaling properties of
the REM:
\begin{equation}
\sum_{\sigma_{V}\in\Omega\left(m\right)}\exp\,[-\beta J'\left(\sigma_{V}\right)\sqrt{K'\left(m\right)}]\stackrel{d}{=}\sum_{\alpha\leq2^{\,R\left(m\right)}}\exp\,[-b\left(m\right)J^{\alpha}\sqrt{R\left(m\right)/2}]
\end{equation}
where we introduced the rescaled size $R$
\begin{equation}
\frac{R\left(m\right)}{N}:=\frac{\phi\left(m\right)}{\log2},
\end{equation}
the rescaled inverse temperature $b$
\begin{equation}
b\left(m\right):=\beta\delta\sqrt{\frac{\left(1-m^{2}\right)N}{R\left(m\right)}},
\end{equation}
the index $\alpha$ (spanning from $1$ to $2^{R}$) and the set of
i.i.d. Gaussian energies $J^{\alpha}$. We can now compute the partition
function
\begin{equation}
Z\left(b,R\right):=\sum_{\alpha\leq2^{\,R}}\exp\,(-bJ^{\alpha}\sqrt{R/2})
\end{equation}
using the formula for the REM free energy \cite{Mezard_Montanari}:
\begin{equation}
\log Z\left(b,R\right)=\left\{ \begin{array}{l}
\frac{1}{4}b^{2}R+R\log2\\
bR\sqrt{\log2}
\end{array}\ \begin{array}{c}
b\leq2\sqrt{\log2}\\
b>2\sqrt{\log2}
\end{array}\right.
\end{equation}
By substituting in the normalization
\begin{multline}
\int_{-1}^{1}dm\,\exp\left(-N\beta\psi m\right)\sum_{\sigma_{V}\in\Omega\left(m\right)}\exp\,[-\beta J'\left(\sigma_{V}\right)\sqrt{K'\left(m\right)}]=\\
=\int_{-1}^{1}dm\,\exp\,[-N\beta\psi m+\log\,Z\left(b\left(m\right),\,R\left(m\right)\right)].\label{eq:xcvzvdcz}
\end{multline}
we find that the correct function to minimize is 
\begin{equation}
\mathcal{F}\left(m\right):=N\beta\psi m-\log Z\left(b\left(m\right),\,R\left(m\right)\right)
\end{equation}
and the function $m_{0}$ is therefore given by the formula
\begin{equation}
m_{0}:=\arg\min\,\left\{ \mathcal{F}\left(m\right)\right\} .
\end{equation}
We now use the usual gradient method to solve this variational problem.
Let's calculate the various components of the REM free energy:
\begin{equation}
bR/N=\beta\delta\sqrt{\frac{\left(1-m^{2}\right)\phi\left(m\right)}{\log2}},\ \ \ b^{2}R/N=\beta{}^{2}\delta^{2}\left(1-m^{2}\right).
\end{equation}
Substituting we obtain the function to minimize:
\begin{equation}
\mathcal{F}\left(m\right)=\left\{ \begin{array}{l}
N\beta\psi m-\frac{N}{4}\beta{}^{2}\delta^{2}\left(1-m^{2}\right)-N\phi\left(m\right)\\
N\beta\psi m-N\beta\delta\sqrt{\left(1-m^{2}\right)\phi\left(m\right)}
\end{array}\ \begin{array}{c}
m^{2}\geq m_{c}^{2}\\
m^{2}<m_{c}^{2}
\end{array}\right.\label{eq:zaza}
\end{equation}
Notice that the first equation is proportional to the arithmetic mean
between the binary entropy $\phi$ and the REM correction, while the
second is proportional to their geometric mean. The critical magnetization
$m_{c}^{2}$ is found through the equation
\begin{equation}
\phi\left(m_{c}\right)=\frac{\beta^{2}\delta^{2}}{4}{\textstyle \left(1-m_{c}^{2}\right)}\label{eq:critical}
\end{equation}
that has two symmetric nontrivial solutions with $m_{c}^{2}>0$ for
$\beta\geq\beta_{c}$ 
\begin{equation}
\beta_{c}:=\frac{2\sqrt{\log2}}{\delta}.
\end{equation}
Otherwise, for $\beta<\beta_{c}$ the first expression of the Eq.
(\ref{eq:zaza}) holds for any magnetization value (i.e., $m_{c}^{2}=0$).
Let us compute the derivative of $\phi$:
\begin{equation}
\partial_{m}\phi\left(m\right):=-\tanh^{-1}\left(m\right)
\end{equation}
Performing also the other computations we find
\begin{equation}
\partial_{m}\mathcal{F}\left(m\right)=\left\{ \begin{array}{l}
N\beta\psi+\frac{N}{2}\beta{}^{2}\delta^{2}m+N\tanh^{-1}\left(m\right)\\
N\beta\psi+N\beta\delta\,m\sqrt{\frac{\phi\left(m\right)}{1-m^{2}}}+\frac{N}{2}\beta\delta\tanh^{-1}\left(m\right)\sqrt{\frac{1-m^{2}}{\phi\left(m\right)}}
\end{array}\ \begin{array}{c}
m^{2}\geq m_{c}^{2}\\
m^{2}<m_{c}^{2}
\end{array}\right.
\end{equation}
Now we can impose the stationary condition:
\begin{equation}
\partial_{m}\mathcal{F}\left(m\right)=0.
\end{equation}
The first equation holds when $m_{0}^{2}$ is above the critical value
$m_{c}^{2}$
\begin{equation}
\tanh^{-1}\left(m_{0}\right)+\frac{\beta^{2}\delta^{2}}{2}\,m_{0}=-\beta\psi,\label{eq:dfgdf}
\end{equation}
the second applies when it is below $m_{c}^{2}$ 
\begin{equation}
\tanh^{-1}\left(m_{0}\right)\sqrt{\frac{1-m_{0}^{2}}{\phi\left(m_{0}\right)}}+2m_{0}\,\sqrt{\frac{\phi\left(m_{0}\right)}{1-m_{0}^{2}}}=-\frac{2\psi}{\delta}.\label{eq:dfgdf-1}
\end{equation}
Notice, the minus sign on the right hand sides is reflecting our sign
convention, where the magnetizations of the ground state are opposite
with respect to the field directions. Since in both Eq.s (\ref{eq:dfgdf}),
(\ref{eq:dfgdf-1}) the left hand sides are odd functions, if we are
interested only in the absolute values of $m_{0}$, then we could
turn them to positive and compute the correct expressions anyway.
Finally, notice that for a vanishing $\beta\delta$ the inverse of
Eq. (\ref{eq:dfgdf}) actually converges to the hyperbolic tangent:
this was the limit considered in \cite{Franchini2023} (Lemma 12,
Section 5).

\section{Acknowledgments}

I would like to thank Giampiero Bardella and Riccardo Balzan (Sapienza
Università di Roma), Pan Liming (USTC) and Giorgio Parisi (Accademia
Nazionale dei Lincei) for interesting discussions. I would also like
to thank an anonymous Referee, for noticing an error in Eq. (\ref{eq:renorma})
and another anonymous Referee, for bringing to my attention ref. \cite{Mezard_Lucibello}.
This project has been partially funded by the European Research Council
(ERC), under the European Union\textquoteright s Horizon 2020 research
and innovation programme (grant agreement Num. {[}694925{]}). Finally,
I'm grateful to Francesco Concetti (UniDistance Suisse) for spotting
the wrong $m_{0}$ in the published version.

\end{document}